\documentclass[11pt]{article}
\usepackage{geometry}                
\usepackage{graphicx}
\usepackage{amssymb}
\usepackage{epstopdf}
\title{EMC Effect, Short-Range Correlations in Nuclei and Neutron Stars}
\author{Mark Strikman\\ 104 Davey Lab, Penn State University, University Park, PA 16802, U.S.A.}

\begin{document}
\maketitle

\begin{abstract}
 The recent $x> 1$ (e,e')  and correlation experiments at 
 momentum transfer $Q^2 \ge 2 \,\mbox{GeV}^2$ confirm presence of  short-range correlations (SRC) in nuclei mostly build of nucleons. 
 Recently we evaluated in a model independent way the dominant photon contribution to the nuclear structure. Taking into account this effect and  using   definition of x consistent with the exact kinematics of eA scattering (with exact sum rules) results in the significant reduction of 
 $R_A(x,Q^2)=F_{2A}(x,Q^2)/F_{2N}(x,Q^2)$ ratio which explains  $\sim $ 50\% of the EMC effect for $x\le$ 0.55 where Fermi motion effects are small. The remaining part of the EMC effect at $x\ge 0.5$ is consistent with dominance of the contribution of SRCs. Implications for extraction of the $F_{2n}/F_{2p}$ ratio are discussed.  Smallness of the non-nucleonic degrees of freedom in nuclei matches well the recent observation of a two-solar mass neutron star, and while large pn SRCs lead to enhancement of the neutron star cooling rate for kT$\le$ 0.01 MeV.
 \end{abstract}

\maketitle

\section{Introduction}
To resolve microscopic structure of nuclei one needs to use high energy high momentum transfer probes. Otherwise the high frequency components of the  nuclear  wave function enter only as  renormalization/cutoff parameters 	in the descriptions of the low energy phenomena, like for example   in chiral effective field theory.

The key questions which can be addressed by using high energy processes and which are relevant for the description of high density cold nuclear matter at the neutron star densities  are (i) can nucleon be good quasiparticles for description of high energy processes off nuclei, (ii) does the notion of the momentum distributions in nuclei make sense for $k\ge m_{\pi}$, (iii) what is probability and structure of the short-range/ high momentum  correlations in nuclei, (iv) what are the most important non-nucleonic degrees of freedom in nuclei, and (v) what is the microscopic origin of intermediate and short-range nuclear forces.
Below we summarize the recent progress in the studies of hard nuclear processes which allows to address several of these questions. 

\section{Recent progress in the studies of the SRCs in nuclei}

Singular nature of $NN$ interaction at large momenta/ small internucleon distances leads to universal structure of SRC and  the  prediction of the scaling of the ratios of the cross sections of $x  > 1$ scattering at sufficiently large $Q^2 \ge 2 \mbox{GeV}^2$ \cite{Frankfurt:1981mk}. In particular for $1+k_F/m_N < x < 2$: 
\begin{equation}
R_A(x,Q^2)=
2\sigma(eA \to e + X)/ A\sigma(e^2H \to e + X).
\end{equation}
Here $a_2(A)$ has the  meaning of the relative probability of the two nucleon SRCs per nucleon in a nucleus and in the deuteron.
The first evidence for such scaling of the ratios was reported in \cite{Frankfurt:1988nt}.  The extensive studies were performed using various data taken at SLAC in \cite{Frankfurt:1993sp}. 
The experiments performed at Jlab allowed to explore the scaling of ratios in the same experiment. 
In \cite{Egiyan:2003vg,Egiyan:2005hs} the scaling relative to $^3$He was established. Very recently the results of the extensive study of the  nucleus/deuteron ratios were reported in \cite{Fomin:2011ng} allowing a high precision determination of the relative probability of the two nucleon SRCs in nuclei and the deuteron.
The results of \cite{Fomin:2011ng} are in a good agreement with the early analysis of \cite{Frankfurt:1993sp}, see Fig.~1.

Several theoretical observations are important for interpretation of the scaling ratios: (a) The invariant energy of the produced system for the interaction off the deuteron is small - $W-m_{^2H} \le \mbox{250 MeV}$ so production of inelastic final states is strongly suppressed. Correspondingly, scattering off  exotic configurations like hidden color configurations which decay into excited baryon states, $\Delta$'s, etc is strongly suppressed in the discussed kinematics, (b) The closure is valid for the final state interaction of  the nucleons of the SRC and the residual nucleus system. Only the f.s.i. between the nucleons of the SRC contributes to the total (e,e') cross section \cite{Frankfurt:1993sp,Frankfurt:2008zv}. Since this interaction is the same for light and heavy nuclei it does not modify the scaling of the ratios, (c) In the limit of large $Q^2$ the cross section is expressed through the light - cone projection of the nuclear density matrix, $\rho_A^N(\alpha)$ that is the integral over all components of the interacting nucleon four momentum except $\alpha\equiv p_-/(m_A/A)$ where $p_- =p_0-(\vec{p}\cdot \vec{q})/ |\vec{q}|$. The ratio of the cross sections reaches a plateau at $x(Q^2)$ corresponding to the scattering off a nucleon with minimal momentum $\sim k_F$ indicating that the dominance of two nucleon SRC sets in just above the Fermi surface. A further confirmation of dominance of two nucleon correlation comes from the observation \cite{Frankfurt:1993sp} of precocious scaling of the ratios plotted as a function of the minimal $\alpha$ for the scattering off two nucleon SRC at rest (the Fermi motion of the pair practically cancels out in such a ratio)\cite{Frankfurt:1988nt}: $\alpha_{tn}= 2 -(q_0-q_3/ 2 m_N)(1+ (\sqrt{W^2-4m_N^2}\/ W)$, where $W^2=4m_N^2+4 q_0m_N-Q^2$.
The precocious  $\alpha_{tn} $ scaling indicates that $R_A$ is equal to  the ratio of the light cone density matrices of the deuteron and nucleus. It also strongly indicates that  SRCs of baryon charge two are predominantly build of two nucleons rather than some exotic states.

\begin{figure}
  \includegraphics[height=.3\textheight]{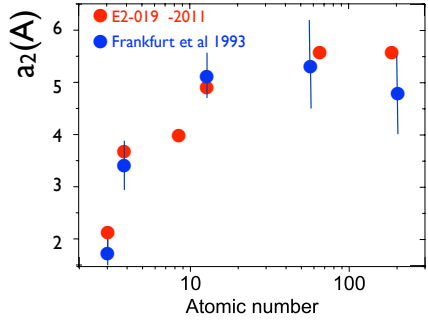}
  \caption{Comparison of the first determination of $a_2(A)$ based on the analysis of the SLAC data \cite{Frankfurt:1993sp} with the most recent Jlab measurements \cite{Fomin:2011ng}.  }
\end{figure}

To probe directly the structure  of the SRCs it is advantageous to study a decay  of SRC after one nucleon of the SRC is removed which is described by the nuclear decay function \cite{Frankfurt:1981mk,Frankfurt:1988nt}. In the two nucleon SRC approximation the decay function is simply expressed through the density matrix as the removal of one of the nucleons of the correlation results in the release of the second nucleon with probability of one.  A series of the experiments was performed at BNL and Jlab which studied (p,2p), (e,e'p)  reactions in the kinematics where a fast proton of the nucleus is knocked out, see review and references in \cite{Subedi:2008zz,Frankfurt:2008zv}. In spite of very different kinematics -- removal of forward moving nucleon in the $^{12}$(p,2p) case and backward moving proton in the $^{12}$C(e,e'p) case, different probes and different momentum transfer $-t \approx \mbox{5 GeV}^2 $ and $Q^2=\mbox{2 GeV}^2$ -- the same of the neutron emission pattern is observed  - the neutron is emitted with a probability $\sim 90\%$ in the direction approximately opposite to the initial proton direction with the correlation setting in very close to $k_F(C) \sim $ 220 MeV/c. The Jlab experiment  observed in the same kinematics both proton and neutron emission in coincidence 
with $e'p $ and found the probability of the proton emission to be about 1/9 of the neutron probability. Hence the data confirm the our theoretical expectation that removal of a fast nucleon is practically always associated 
with the emission of the nucleon in the opposite direction with the SRC contribution providing the dominant component of the nuclear wave function starting close to the Fermi momentum. The large pn/pp ratio also confirms  the standard expectation of the nuclear physics that short-range interactions are much stronger in the isospin zero channel than in the isospin one channel. Saturation of the probability provides an independent confirmation of the conclusion that at least up to momenta $ \sim 500 \div 600 $ MeV/c SRC predominantly consist of two nucleons.

\section{New developments in the studies of the EMC effect}

The deep inelastic scattering off nuclei can be described in the impulse approximation as the  convolution of the LC density matrix and the elementary cross section:
\begin{equation}
F_{2A}(x, Q^2) = \int_0^A {d\alpha\over \alpha} \rho_A^N(\alpha) F_{2N}({x\over \alpha}, Q^2),
\label{conv}
\end{equation} 
where $x=AQ^2/2q_0m_A$ and 
 $\rho_A^N(\alpha))$ satisfies the baryon charge conservation sum rule:
 $\int_0^A {d\alpha\over \alpha} \rho_A^N(\alpha)=A$. If the nucleus in the fast frame consists only of nucleons,  $\rho_A^N(\alpha))$ also satisfies the momentum sum rule: 
$\int_0^A \alpha{d\alpha\over \alpha} \rho_A^N(\alpha)=A$. Together these sum rules imply that in the many nucleon approximation the EMC ratio $R_A(x,Q^2)= F_{2A}(x,Q^2) /F_{2N}(x,Q^2)$ should be  slightly below one for a range of x below $x_0=2/(1+n)$ where $F_{2N}(x)\propto (1-x)^n$ and exceed one for $x> x_0$. Significant deviation of the EMC ratio from these expectations clearly indicates presence of non-nucleonic degrees of freedom in nuclei.

We have demonstrated recently  \cite{Frankfurt:2010cb}   that two effects should be taken into account before considering modifications of the many nucleon approximation for the nuclear wave function: presence of the Coulomb field in a fast nucleon  and the difference between the definition of the Bjorken variable in the theoretical expression (Eq.~1) - $x=AQ^2/2q_0m_A$,  and the one used in the experimental papers -
$x_p=Q^2/2q_0m_p$.

Atomic nuclei carry electric charge. Therefore the Coulomb field of a nucleus is a
fundamental property of the nucleus in its rest frame. Under the  Lorentz transformation 
to the frame where the nucleus  has a large momentum, the   rest frame nucleus Coulomb field  is 
transformed into the field of equivalent photons. This phenomenon is well known as 
Fermi - Weizsacker - Williams  approximation for  the wave function of a rapid projectile with 
nonzero electric charge. Application of this techique  allows  to  evaluate  the role of photon degrees of freedom 
in the partonic nucleus structure. In particular we find for an additional 
(to the case of the system of free nucleons)
light-cone (LC)  fraction of the nucleus momentum carried by photons \cite{Frankfurt:2010cb} \footnote{This formulae corrects corresponding expression of ref.\cite{Frankfurt:2010cb} where numerical coefficient was overestimated by a factor $\sim $ 2.6.}
\begin{equation}
 \lambda_{\gamma}=\int_0^1\, dx xP_{\gamma}(x,Q^2)={\alpha_{em}} {2\over \sqrt{3\pi}} {Z(Z-1)\over A}{1 \over m_NR_A}.
 \label{ww7}
 \end{equation}
 The leading effect $\propto Z^2$ is due to the  coherent 
emission by the nucleus as a whole, and the term $\propto Z$ is due to the subtraction of the incoherent  emission of photons by individual protons. Numerically   $\lambda_{\gamma}(^{12}C) = .11\% ; \lambda_{\gamma}(^{56}Fe) =.35\%; \lambda_{\gamma}(^{197}Au) =0.65\%$. 
Presence of the dynamic photon field  modifies the parton momentum sum rule:
\begin{equation}
 \int_0^A \left[(1/A)(xV_{A}(x, Q^2) +xS_{A}(x, Q^2) +xG_{A}(x,Q^2))\right]dx=1- \lambda_{\gamma}.
 \label{sumrule}
 \end{equation}
 This effect  has to be taken into account in the analyses of the nuclear pdfs. In particular it leads to a $\sim 1.3 \% $ reduction of the momentum fraction carried by gluons in heavy nuclei since it is determined using Eq.\ref{sumrule} and the $F_{2A}/F_{2^2H}$ data.

The depletion of the LC fraction carried by nucleons leads to a significant EMC  effect for $A\ge 50$:
\begin{equation}
R_A(x,Q^2) -1 = - \lambda_{\gamma}xF_{N}^{\prime}(x,Q^2)/ F_{N}(x,Q^2).
 \label{RA}
 \end{equation}
  \begin{figure}
  \includegraphics[height=.35\textheight]{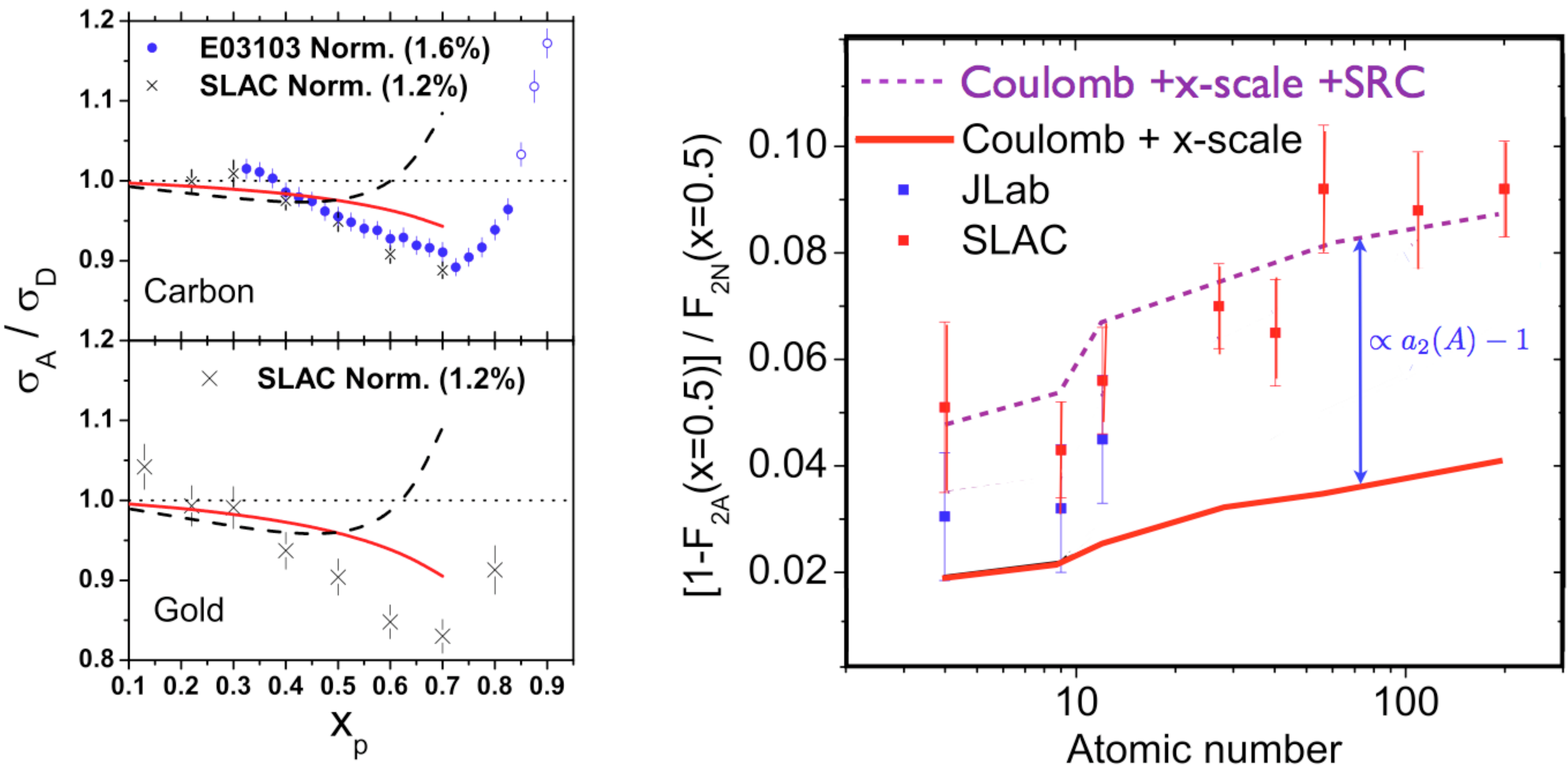}
   \caption{Solid lines are results of calculation taking into account the Coulomb effect and the effect of proper definition of x. In Fig.~2b the dashed line is the contribution of the hadronic  EMC effect due to SRCs normalized for large A with the  A-dependence  given by $a_2(A)$ from \cite{Fomin:2011ng}. The dashes lines in Fig.~2b are result of adding the effect of the Fermi motion.
  The data are from the SLAC and Jlab experiments \cite{Gomez:1993ri,Seely:2009gt}. }
\end{figure}
 Correcting for the difference between $x_p$ used in the experimental papers and $x$ which enters in the convolution expression Eq.\ref{conv} \cite{Frankfurt:1981mk} also leads to a EMC like effect for $R_A$.
 It can be taken into account by the substitution
 $\lambda_{\gamma} \to \lambda_{\gamma} + (\epsilon_A- \epsilon_{^2H}  - (m_n-m_p) (N-Z)/A)/m_p$
 in Eq.~\ref{RA}.
  The Coulomb field  effect is much smaller  than the x-rescaling  for $A\le 12$,  while for $A\sim 200$ it is  as large as the x-rescaling effect.  Combined these two effects leads to the solid curves in Fig.~2. One can see that  these two {\it model independent}  effects explain $\approx 50\%$ of the EMC effect for $x\le 0.5$ where Fermi motion effects are small. For  $x > 0.5$ where Fermi motion contribution becomes large, an additional effect of modification  of the hadronic component of the nucleus wave function is necessary mostly  to compensate   the Fermi motion effect. The "hadronic EMC effect" is $\approx$ 4\% for $A\ge 50$ for x=0.5 and  grows rapidly with further increase of x: $\sim $ 15\% for x=0.6, $\sim$ 25\% for x=0.7. This steep  $x$-dependence
  is  consistent 
   with the expectation of the color screening model that  maximal suppression $\sim 20\%$ occurs for very large $x$ where point like configurations dominate in $F_{2N}$   \cite{Frankfurt:1985cv,Ciofi}.

 It was demonstrated in  \cite{Frankfurt:1985cv,Ciofi} that a  bound nucleon deformation is proportional to the nucleon's  kinetic energy (the nucleon off-shellness). Hence the EMC effect is proportional to the average kinetic nucleon energy, which  is dominated by the contribution of the SRCs.
 In Fig.~2b we plot $1-R_A(x=0.5)$ since the Fermi motion does not contribute for this x \cite{Frankfurt:1981mk}. One can see that the A-dependence of the "extra" EMC effect for $F_{2A}/F_{2^2H}$ is indeed roughly consistent with the measured A-dependence of $a_2(A)-1$ (the same is true for x=0.6, 0.7).

 Our analysis indicates that the non-nucleonic components contribute significantly only in nucleons with  $x\ge 0.5$ quarks. Such configurations   occurs with a very small probability $\sim 2\%$. Hence we conclude that the probability of exotic component  
relevant for the large x  EMC effect is $\sim  0.2\%$. Since the residual effect for smaller $x$ is $\le 1\div 2 \%$ we conclude that overall the  probability of the exotic component in nuclei is $\le 2\%$. This is consistent with the results of the analysis described in Sect. 2 that  SRCs are dominated by the nucleonic degrees of freedom.
 
  In the case of the scattering  off the deuteron the Coulomb and x-rescaling effects are practically negligible and  only hadronic effect is present. Since the hadronic EMC effect is proportional to the average nucleon kinetic energy (average virtuality) it is expected to be approximately factor of 4 smaller for the deuteron than for medium and heavy nuclei \cite{Frankfurt:1985cv}, \cite{Ciofi}.
   As a result the EMC effect for the deuteron ($R_D(x,Q^2)=F_{2D}/F_{2N}(x,Q^2)$) say for $x=0.5$ is  approximately 1/4 of  the difference between the   dashed and solid curves in Fig.~2b for $A\ge 50$ - that is 
  $1 - R_D(0.5,Q^2) \approx  0.01$ (which is 
   a factor of $\sim 2$ smaller than if one assumes that all the EMC effect is due to the scattering off the SRCs) leading to a reduction of the  extracted $F_{2n}/F_{2p}$ ratio at large $x$.
  
  \section{Some implications for neutron stars}
  The small probability of the nonnucleonic degrees of freedom in nuclei including SRC which follows from the studies of the hard nuclear phenomena fits well with the  recent observation\cite{Demorest:2010bx}  of a heavy neutron star of about two Solar masses  - models where nonnucleonic degrees of freedom are easily excited  do not allow existence of such heavy neutron stars.
  
   Our focus  is on the outer core where nucleon density is close to the  nuclear one: $\rho \sim (2\div 3) \rho_0$, where $\rho_0 \approx$ 0.16 nucleon/fm$^3$ and the ratio of the proton and neutron densities $x\sim 1/10$, corresponding to 
  \begin{equation}
k_F(p)/k_F(n) = (N_p/N_n)^{1/3}\equiv x^{1/3} \ll 1. 
\end{equation}
Since the probability of the pn SRC grows with the neutron density which  is a factor of $4 \div 6$ higher for $\rho \sim (2\div 3) \rho_0$. As a result 
the neutron gas "heats" the  proton gas leading to practical disappearance of the proton Fermi surface \cite{Frankfurt:2008zv}.

The high momentum tail of proton, neutron distributions are directly calculable. In  the leading order in $k_F^2/k^2$ the occupation numbers for protons and neutrons with momenta above Fermi surface are
\begin{eqnarray}
f_{n}(k,T=0) \approx \left({\rho_{n}}\right)^2(\left(\frac{V_{nn}(k)}{k^2/m_N}\right)^2+ 2x\left(\frac{V_{pn}(k)}{k^2/m_N}\right)^2), \\ \nonumber 
\, f_{p}(k,T=0) \approx  \left({\rho_{n}}\right)^2(x^2 \left(\frac{V_{pp}(k)}{k^2/m_N}\right)^2 +
2x \left(\frac{V_{pn}(k)}{k^2/m_N}\right)^2).
\end{eqnarray}
Since there is an equal number of protons and neutrons   above Fermi surface, but $x\ll 1$,  the 
 effect is much larger for protons than for  neutrons.

As a result, the internucleon interaction tends to equilibrate momenta of protons and neutrons -strong departure from the ideal gas approximation. The Migdal jump in the proton momentum distribution almost disappears in this limit. The suppression of the proton Fermi surface leads to the suppression of the proton superconductivity. At the same time the superfluidity of neutrons and proton-neutron pairs is not excluded.

Another effect is the large enhancement of neutrino cooling of the neutron stars at finite temperatures \cite{Frankfurt:2008zv}.
The enhancement (factor of R as compared to the URCA process)  is due to presence of the proton holes in the proton Fermi sea.  For example taking $x =0.1$, and the neutron density ~ $ \rho_0$ ,   we find  for the temperature kT $\ll$  1 MeV:
\begin{equation}
R \approx   0.1 (MeV/kT)^{3/2},
\end{equation}
and much larger enhancement for $x \ll 0.1$ where the URCA process is not effective.
Since the temperature of the isolated neutron star  drop below 0.01 MeV after one year, the discussed mechanism leads to a   large enhancement
of cooling.
\section{Conclusions}
The impressive experimental progress of the last few years - discovery of strong short range correlations in nuclei with strong dominance of I=0 SRC -  confirmed a series of our predictions of 80's  and  has proven validity of general strategy of using hard nuclear reactions for probing microscopic nuclear structure. It provides a solid basis for the further studies. Several experiments are under way and several are already a part of the planned  12 GeV Jlab research. The hadronic EMC effect is a factor $\sim 2$ smaller for $x \le  0.5$ than was thought previously,
 but it kicks in  rapidly at $x >  0.5$ implying that the tagged structure function studies should observe a transition from nearly free nucleon like $F_{2N}$ for $x\le 0.45$  to a strongly deformed $F_{2N}$ at $x\sim 0.6$. Nucleons remain practically undeformed  up to the local densities comparable to the neutron star densities which is  consistent with a stiff equation of state for the neutron stars.
A direct observation of  3N SRCs, nonnucleonic degrees of freedom in nuclei ($\Delta$-isobar like configurations, etc) which are of direct
 relevance for the neutron stars core dynamics are on the top of the agenda for the future research. Observation of these effects will be   one  of the aims of our data mining program at Jlab, as well as of a number of experiments at 12 GeV.
 Complementary experiments with hadron beams (FAIR,  J-PARC) are highly desirable.

\end{document}